\documentclass{aa}  

\usepackage[switch]{lineno}
\usepackage{graphicx}	% Including figure files
\usepackage{amsmath}	% Advanced maths commands
\usepackage{amssymb}	% Extra maths symbols%%%%%%%%%%%%%%%%%%%%%%%%%%%%%%%%%%%%%%%%
\usepackage{txfonts}
\usepackage{hyperref}
\usepackage{xcolor}
\usepackage[utf8]{inputenc}
\usepackage[normalem]{ulem}
\usepackage{amsmath}

\begin{document}

   \title{Cosmic Ray bubbles from nova super remnants and their contribution to local cosmic ray spectra.}

   %\subtitle{I. Overviewing the $\kappa$-mechanism}

   \author{Rub\'en L\'opez-Coto\thanks{E-mail: rlopezcoto@iaa.es}
          \inst{1}
          \and
          David Green\inst{2}
          \and
          Javier M\'endez-Gallego\inst{1}
          \and
          Emma de O\~na Wilhelmi\inst{3}
          \fnmsep
          }

   \institute{Instituto de Astrof\'isica de Andaluc\'ia, CSIC, Granada, 18008, Spain
              %\email{wuchterl@amok.ast.univie.ac.at}
         \and
Max-Planck-Institut f\"ur Physik, D-80805 M\"unchen, Germany
\and
Deutsches Elektronen-Synchrotron DESY, Platanenallee 6, 15738 Zeuthen, Germany
             }

   \date{Received xx; accepted yy}

% Several new phenomena have been surrounding the area of study of the repeating thermonuclear explosions called novae. For example, recurrent novae have been proven to be efficient cosmic ray hadronic accelerators thanks to the recent observations of RS Ophiuchi (RS Oph) by different $\gamma$-ray instruments. They have also demonstrated to have the ability to carve large cavities into the Interstellar Medium with parallelisms with the remnants of supernovae. Due to the energy budget of the accelerated protons in these objects and the frequency of nova explosions in the Milky Way, this opens the question of whether novae could significantly contribute to the energy budget of cosmic rays in the Milky Way. We study in this paper the percentage covered in our Galaxy, the contribution to local cosmic ray fluxes and we also use cosmic rays to put strong limits on the density of the region carved by these nova remnants using the example of RS Oph.

% \abstract{}{}{}{}{} 
% 5 {} token are mandatory
 
  \abstract
   {Several new phenomena have been surrounding the area of study of the repeating thermonuclear explosions called novae. For example, recurrent novae have been proven to be efficient cosmic ray hadronic accelerators thanks to the recent observations of RS Ophiuchi by different $\gamma$-ray instruments. Novae have also demonstrated to have the ability to carve large cavities into the Interstellar Medium with parallelisms with the remnants of supernovae.
}
   {Calculate what is the effect of novae in their surrounding media and to which distances these effects dominate over the average quantities that are measured in the ISM. }
   {We calculate the filling factor of novae and their contribution to cosmic ray fluxes using cosmic ray propagation codes. To limit what is the atomic density of the Interstellar Medium (ISM) surrounding the region around RS Oph, we use {\it Fermi}-LAT observations of the region.}
   {The filling factor of novae in the Galaxy is not significant under all assumptions done in the paper. They do not dominate over the local cosmic ray fluxes, even at the lowest energies, for distances larger than a few parsec. The particle density of the ISM surrounding them is, however, very much modified, lowering it more than one order of magnitude with respect to galactic averages, confirming estimates done using other observatories.}
   {Even though at global galactic distances, novae do not seem to be dominating cosmic ray transport, they have the power to modify the conditions of their surrounding ISM over parsec distances.}


   \maketitle
%
%-------------------------------------------------------------------
%1/ introducion
%* nova as cr accelerator
%*  impulsive and continuous accelerators and paper on the super shell to justify the second
%2/ Analysis: 
%impulsive -> de first part
%continuous -> the second part
%3/ ISM, low density, etc

\section{Introduction}
%nova general
Cataclysmic variable stars are semi-detached binaries that consist of a primary White Dwarf (WD) accreting material from a companion star. Novae constitute a subclass of cataclysmic variable stars and they consist of outbursts caused by the accumulation of hydrogen-rich material on the surface of the WD which results in a thermonuclear runaway creating thermonuclear explosions and resulting fast shocks. A nova is denominated as classical when only one outburst has been reported and recurrent if they have documented repeated eruptions \citep{classical_novae}. The repetition period can vary from years to decades or almost centuries \citep{recurrent_novae, recurrent_novae_Darnley}. The further sub-classification of symbiotic denotes that the donor star has evolved from the main sequence such as a red giant star and therefore the WD is immersed in the donor star's wind.  Novae can increase their brightness from 6 up to 19 magnitudes during these outbursts, which can last from weeks to months, see \citep{classical_novae} for a review.
%nova high energy
Novae have been studied from radio up to X-rays for decades. It was only 14 years ago that there was the first report of High Energy (HE; $E>$100 MeV) $\gamma$-ray emission from the recurrent symbiotic nova V407 Cyg by {\it Fermi}-LAT \citep{2010Sci...329..817A}. Later on, classical novae were also established as HE gamma-ray emitters \citep{2014Sci...345..554A} and the number of novae detected in HE gamma rays has been increasing since then \citep{2016ApJ...826..142C, 2018A&A...609A.120F, 2020NatAs...4..776A}. Even though these observations point toward the acceleration of Cosmic Rays (CRs) by novae, the origin of this gamma-ray emission was still unclear. Their spectra could be well-represented by models considering leptons producing gamma rays by inverse Compton up-scattering photons from the photosphere or protons decaying into $\pi^0$ and producing gamma rays \citep{2012PhRvD..86f3011S}.
%nova very high energy
RS Ophiuchi (RS Oph) is a recurrent symbiotic nova that displays major outbursts every 15-20 years \citep{1994AJ....108.2259D}. Its last outburst took place on August 8th, 2021 and the event was panchromatically observed and is one of the best studied transient nova events. It was followed up by observing energies ranging from radio up to Very-High-Energy (VHE; $E>$100 GeV) gamma rays. In the HE regime, \emph{Fermi}-LAT measured gamma rays coming from the direction of RS Oph \citep{Fermi_RSOph,HESS_RSOph}. Most importantly, at VHE gamma rays, RS Oph was also detected by Cherenkov telescopes in the TeV regime, including MAGIC \citep{MAGIC_RSOph} and the LST-1 of CTAO LST Collaboration \citep{rsoph_lst} in the Northern hemisphere and H.E.S.S. \citep{HESS_RSOph} in the Southern one. 
In X-rays, it was observed by {\em Chandra} and {\em XMM-Newton} \citep{2021ATel14906....1O}, {\em INTEGRAL}, \citep{2021ATel14855....1F}, {\em MAXI}\citep{2021ATel14846....1S}, {\em NICER}\citep{2021ATel14850....1E}, {\em NuSTAR}\citep{2021ATel14872....1L} and {\em Swift} \citep{2022MNRAS.514.1557P}, in the optical, photometry and spectroscopic observations were also carried out \citep{2021ATel14858....1T, 2021ATel14895....1M, 2021ATel14840....1M, 2021ATel14860....1M, 2021ATel14838....1T, 2021ATel14863....1N,  2021ATel14868....1S, 2021ATel14881....1S, 2021ATel14883....1S, 2021ATel14909....1F, 2021ATel14972....1R, 2021ATel14974....1Z, 2022ATel15330....1Z}
, together with infra-red \citep{2021ATel14866....1W} and radio \citep{2021ATel14886....1S, 2021ATel14908....1P, 2021ATel14849....1W} observations. Upper limits on the neutrino flux were also established by IceCube Multimessenger  \citep{2021ATel14851....1P}.
In the gamma-ray energy range, the continuation of the spectrum and similar flux decay between the HE and VHE measurements point to a common radiation component arising from a single high-speed shock (a two shock model is presented in \citet{Diesing_2023}). Independent of the multiplicity of the shocks in which particles are accelerated, the current consensus is that the underlying particle population producing this emission is of hadronic origin. A pure leptonic origin can still be possible (see \cite{2022MNRAS.515.1644B, 2023JHEAp..38...22B}, but the hadronic interpretation is strengthened by energetic and multi-wavelength temporal evolution arguments (see \citet{MAGIC_RSOph, HESS_RSOph} for details). 
%\wipeouteow{The reasons described in \citet{MAGIC_RSOph} are the easier fitting of the gamma-ray spectra by a proton model with an original energy dependence of $E^{-2}$, a better fitting of the HE part of the spectrum by the hadronic model, a hint of spectral hardening and a similar decay for the optical and HE emission. It should be noted, that a leptonic origin of the gamma-ray emission can also be possible other works that claim that a leptonic origin of the gamma-ray emission is also possible and make predictions for the particle acceleration and persistent emission of recurrent novae \citep{2022MNRAS.515.1644B, 2023JHEAp..38...22B}.}
The efficient acceleration of hadronic CRs in nova shocks, converting a significant fraction of the nova explosion energy into accelerated CRs that reach an energy budget of $\sim10^{43}$ erg, justifies the investigation of a sizeable contribution of novae to the CR sea. \citet{MAGIC_RSOph} and \citet{HESS_RSOph} argue that the contribution of each single outburst (i.e. an impulsive injection of CRs) is sub-dominant to the average Galactic cosmic-ray population beyond $\sim$1\,pc.

However, the contribution of these recurrent novae (RNe), defined as the subclass of the cataclysmic variables that experience repeated thermonuclear eruptions on time scales of a human lifetime, can last millions of years, injecting protons semi-continuously in the surrounding environment during the evolutionary path of the system \citep{2023MNRAS.521.3004H}. Due to these continual eruptions, a dynamical structure, either a nova shell or a larger remnant is formed. This object, called nova super-remnant (NSR) has been recently identified in the Andromeda Galaxy around the nova M31N 2008-12a  \citep{2016ApJ...833..149D}, extending over a size that rivals that of Supernova Remnants (SNRs) (with $\sim$100 pc radius), perhaps more similar to superbubbles observed around massive stellar clusters \citep{2011Sci...334.1103A,2019NatAs...3..561A}. Even though this effect has been simulated and studied leading to an apparent shortage of these systems in the Local Group \citep{2024MNRAS.528.3531H}, a cavity in the far-infrared archival IRAS images fulfilling these characteristics has been found in the region surrounding RS Oph \citep{2024MNRAS.529L.175H}. The cavity is described as an ellipse with a semimajor axis of $\sim$40\,arcmin modeled as the effect of 650,000 eruptions that took place during a period of 25 Myr. This corresponds to a projected size of $\sim$16\,pc, for a distance of 1.4\,kpc \citep{2008ASPC..401...52B}, although the distance to RS Oph has been subject of intense debate (see section C.1 of \citep{MAGIC_RSOph}) and we will also consider a distance of 2.45 kpc as derived in \cite{Rupen_2008} based on VLBA radio imaging of the nova shock expansion. Assuming a similar efficiency in the previous outbursts, one would naively expect an accumulation of CRs being injected continuously in a lifetime of Myrs. Each of these outbursts is expected to excite very turbulent plasma around the nova, that fills up the cavity, potentially confining CRs within the bubble. These CRs can eventually radiate in gamma rays through proton-proton interaction when interacting with the tenuous gas within the cavity, and reach detectable levels if the accumulated CRs are energetic enough.

In this work, we investigate the contribution of RNe to the CRs as a continuous particle accelerator. In Section \ref{sec:sec2} we estimate the contribution of the accumulated protons to the CRs sea, as a function of the distance to the nova and the filling fraction considering RS Oph as a prototype of RNe. In Section \ref{sec:sec3}, we derive upper limits on the NSR gamma-ray emission, which translates to constraints on the proton and density of the Interstellar Medium (ISM) at 16\,pc surrounding RS Oph.

\section{RNe as local and global CR accelerators}
\label{sec:sec2}
%\addtexteow{I dont like contributors, ideas? }\addtextdmg{sources?} \addtextrlc{accelerators?}

If the TeV outburst of RS Oph is to be considered as an archetypal of the acceleration of protons in RNe, it can be shown that the contribution to the CR sea would be no more than 0.2\% \citep{MAGIC_RSOph}. 
Likewise, if we reduce the problem to the local contribution of CRs around the star, we also find that the energy density of the CRs injected by the nova would dominate over the energy density of the CRs in the Milky Way (MW) only for a region with a radius of $R_{\rm nova}\lesssim$0.5 pc, assuming a homogeneous distribution of the CRs in the region \citep{MAGIC_RSOph}. The estimations above concern the single contribution of a given nova. We can now entertain the contribution of a population of similar classical novae independently exploding in the MW to calculate the total filling factor of their regions of influence. 

\subsection{Filling factor of novae in the Milky Way}
\begin{figure}
	\includegraphics[width=\columnwidth]{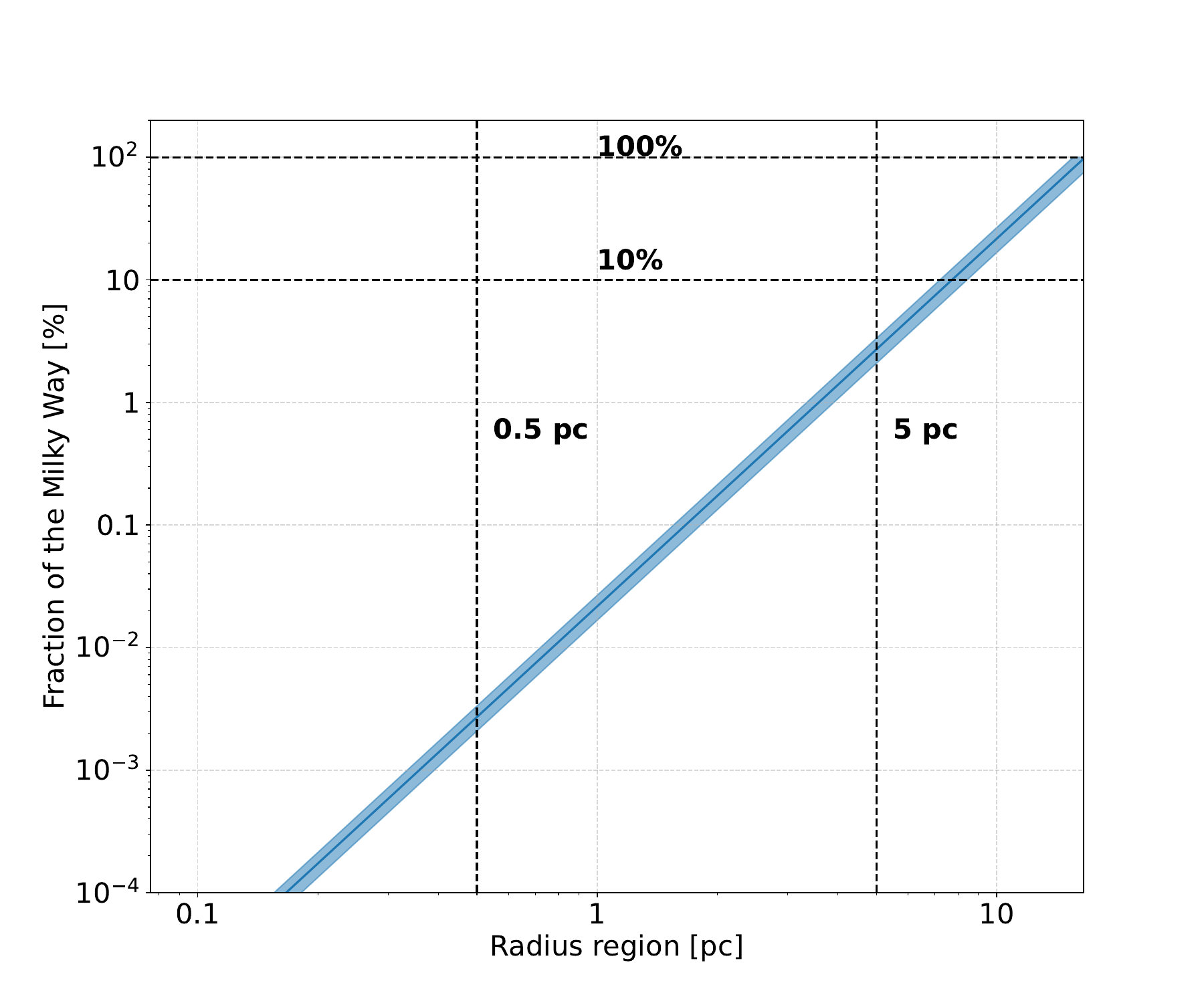}
    \caption{Fraction of the Galaxy covered by CR bubbles created by novae. The band represents the error in the number of novae per year as estimated by \cite{2022ApJ...937...64K}.}
    \label{fig:fraction_galaxy}
\end{figure}

The rate of novae in the MW ($N=50^{+31}_{-23}$ yr$^{-1}$  \citep{2017ApJ...834..196S} or $N=26\pm6$ yr$^{-1}$ in more recent estimates\citep{2022ApJ...937...64K}) makes one wonder if a significant fraction of the MW could be covered by the remnants of these explosions. Without making any initial assumptions about the volume that is covered by a nova remnant, let us compute what would be the percentage of the MW covered by these objects. To estimate the volume of the MW, we consider that it is contained in a cylinder of radius $R_{\rm MW}=20$ kpc and height $H_{\rm MW}=200$ pc \citep{2023Galax..11...77V}. Let us make the very simplistic and conservative assumption that every nova creates a sphere of radius $R_{\mathrm{nova}}$ without considering any overlapping between the regions created by different novae, the fraction (F) covered is therefore given by:

\begin{equation}
\mathrm{F} = N\frac{4/3\pi R_{\rm nova}^3}{\pi R_{\rm MW}^2 2 H_{\rm MW}} t \approx 2\times10^{-4} \left(\frac{R_{\rm nova}}{1 \mathrm{pc}}\right)^3 \left(\frac{t}{10^6 \mathrm{yr}}\right) 
\end{equation}

where $t$ is the time that is considered for nova eruptions to sustain such cavity and we assume $N=26\pm6$ yr$^{-1}$ \citep{2022ApJ...937...64K}.  This is represented in Figure \ref{fig:fraction_galaxy} for the assumption of $t=1$ Myr, together with different significant radii and percentages of the Galaxy covered. We can see that even in the most optimistic scenario of no overlapping between regions, this novae population would cover a significant fraction ($>$10 \% of the MW) only in the case of extending their region of influence for regions with a radius larger than 5 pc and/or lasting for a significant time longer than 1 Myr. Let us remember that, as a first order approximation, the radius of influence over the ISM for a single nova calculated using Equation 16 of \cite{MAGIC_RSOph} is $\sim$0.5 pc and even though that for a RN can reach distances larger than a few parsec (Equation 17 of \cite{MAGIC_RSOph}), the volume of the region of influence of a RN is smaller than that of independent single novae that erupted the same number of times as the recurrent one. 
%for which a fraction of the galaxy $<0.01\%$ of the MW would be covered by nova explosions according to Fig. \ref{fig:fraction_galaxy}. 

%\addtexteow{This is in contradiction with our argument here saying that it should fill up the NSR... I think we need a reorder also here}.
%that is more than one order of magnitude larger than that calculated for single nova explosions in Appendix B.2.1 of \citet{MAGIC_RSOph}. 

%\wipeouteow{I think we need first to put the second part, in which we estimate the 'radius of influence', that is, everything you did with the EDGE code, and then say, okay it is only like 10pc. What now if we have many, in such case, we can estimate the filling factor in the MW. Then we say, we can constrain this even further by looking into the supershell. And end up saying, we will look into the other CV --- I think we should have a look at the other CV, it is just one more... no? } 

%comparing the energy density of CRs in the Milky Way (MW) with that injected by a single nova eruption the energy density of the CRs injected by the nova would dominate that of the interstellar medium for a radius $R_{\rm eruption}\lesssim$0.5 pc, assuming a homogeneous distribution of the surround gas 
%$$, the energy density of the CRs injected by the nova would dominate that of the interstellar medium for a radius $R_{\rm eruption}\lesssim$0.5 pc. 

\subsection{Local CR contribution}
\begin{figure}
	\includegraphics[width=1.1\columnwidth]{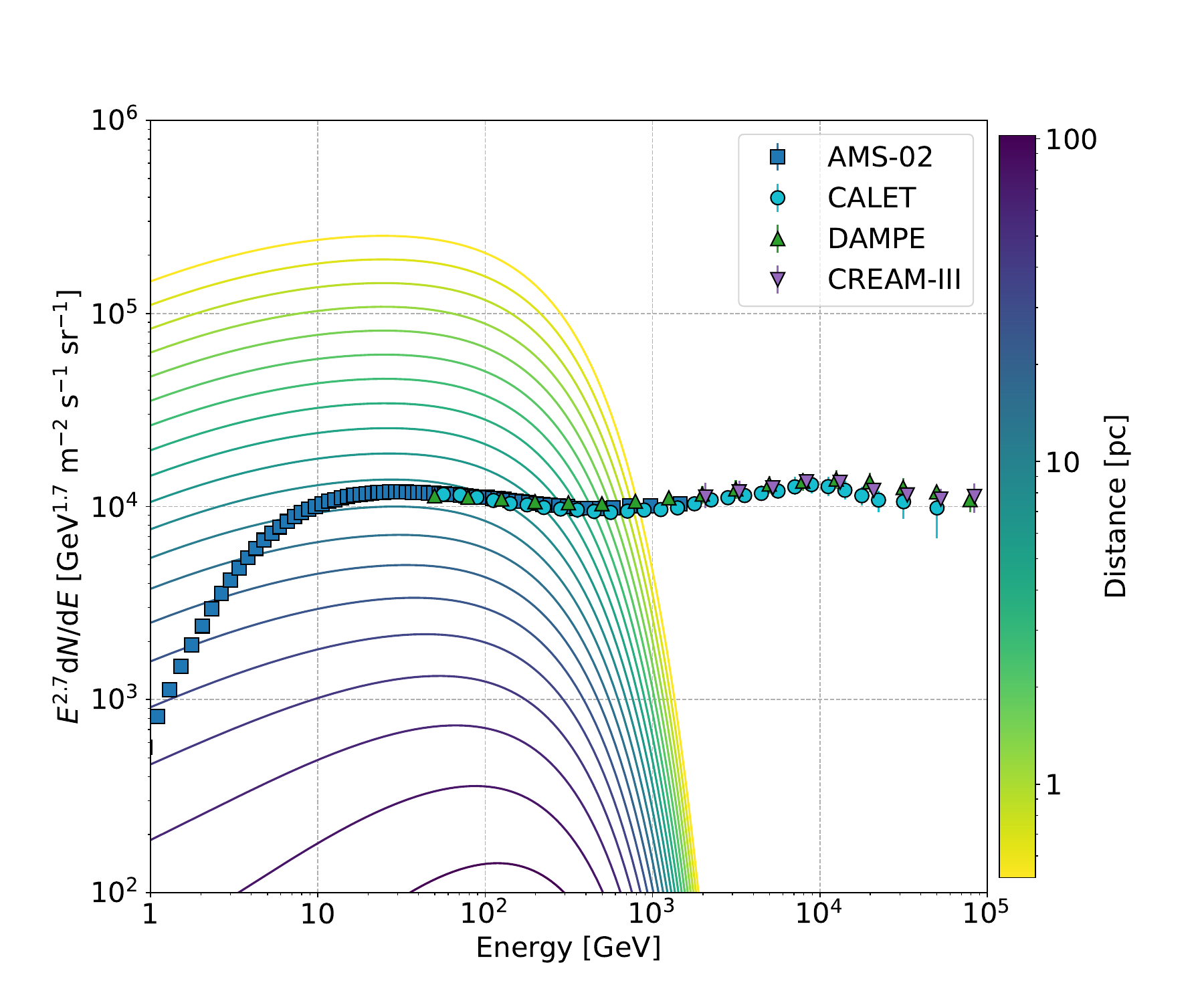}
    \caption{Proton flux produced by a single RN at different distances from the source. For comparison, we show the local proton fluxes locally measured by AMS-02 \citep{2015PhRvL.114q1103A}, CALET \citep{2022PhRvL.129j1102A}, DAMPE \citep{2019SciA....5.3793A} and CREAM-III \citep{2022ApJ...940..107C}.}
    \label{fig:flux_earth_protons}
\end{figure}

Let us go one step further and compute what the contribution of novae to the local CR spectrum surrounding a RN is. For the calculations, we used the EDGE code \citep{2018APh...102....1L} assuming a source continuously injecting protons during 25 Myr. We assumed the optimistic scenario in which the total injected energy in protons per nova eruption is the same as in RS Oph and equal to $4.4 \times 10^{43}$ erg \citep{MAGIC_RSOph} , with the same evolution considered in \citep{2024MNRAS.529L.175H} for a total of 650,000 eruptions during the lifetime considered. CRs are considered to propagate isotropically from the central sources and due to the impulsive behaviour of the injection, we consider them to diffuse away from the central source with an energy-dependent diffusion given by $D(E)=D_0(1+E/E_0)^\delta$ \citep{1995PhRvD..52.3265A} where  $E_0=3$ GeV and $\delta$=0.33 assuming Kolmogorov turbulence and $D_0=1.4\times10^{26}$ cm$^2$/s assuming slow diffusion on a turbulent region as measured in \citep{2017Sci...358..911A} and comparable to that in regions much more turbulent than the ISM \citep{cocoon}. We calculate the CR flux at different distances from the nova explosion and compare them to the one locally measured at Earth in Figure \ref{fig:flux_earth_protons}. Ignoring the region below a few GeV that is affected by solar modulation, we can see that the spectrum produced by a RN only dominates over the local spectrum for a region of $\sim$ 10 pc, similar to what was deducted using simple calculations. We therefore conclude that to obtain a significant contribution of a nova to the CR spectrum measured at Earth, the nova would need to erupt at a distance not further than few tens of parsec with a rate of few decades. Note that we are assuming a slow propagation due to a turbulent region, and for faster propagations, the dominance of the CR spectrum injected by the RN would be reduced with respect to that of the ISM. In case of slower propagation, close to the Bohm limit, or confinement of these CRs, the dominance of NSRs could extend to regions reaching the tens of pc distance, but one would also need to take into account larger cooling effects due to this confinement. 
% \begin{figure}
% 	\includegraphics[width=\columnwidth]{figures/energy_density_vs_distance.pdf}
%     \caption{Energy density vs distance for protons.}
%     \label{fig:energy_density_protons}
% \end{figure}

%\subsection{Leptonic CRs?}

% Even though there are no
% Immediate injection of electrons would clearly lead to . If we consider late time acceleration like that put forward in \citep{}. Even in the very optimistic case of getting a similar amount of energy into late-time electrons as that needed to explain the emission. Comparing the results for 

%H.E.S.S. \cite{HESS_electrons} presented a measurement of the LAES extending up to $\sim$20~TeV with a break at $\sim$900~GeV, later confirmed by DAMPE \cite{Ambrosi17} and CALET \cite{Adriani17}. 

% \begin{figure}
% 	\includegraphics[width=\columnwidth]{figures/distance_vs_energy_electrons.pdf}
%     \caption{Energy density of electrons. }
%     \label{fig:energy_density_electrons}
% \end{figure}

% The flux at Earth for electrons is:
% \begin{figure}
% 	\includegraphics[width=\columnwidth]{figures/flux_earth_electrons.pdf}
%     \caption{Flux at Earth for electrons. Data from AMS \citep{AMS14}, {\it Fermi}-LAT \citep{Ackermann12}, DAMPE \citep{Ambrosi17}, CALET \citep{Adriani17} and H.E.S.S.}
%     \label{fig:flux_earth_electrons}
% \end{figure}

% \section{Other HE novae}
% As discussed in \cite{MAGIC_RSOph}, other novae may be significantly contributing

% Detectability with current generation of IACTs. 

% Conclusion: Novae just need to be either nearby or powerful. Classical novae can also contribute significantly to Galactic CRs.

% Supernovae Ia may be detected

%\begin{figure}
%	\includegraphics[width=\columnwidth]{figures/detectability.pdf}
%    \caption{Detectability by MAGIC.}
%    \label{fig:detectability_magic}
%\end{figure}

\section{Nova super remnants as continuous CR injectors: the case of RS Ophiuchi}
\label{sec:sec3}
In Section \ref{sec:sec2} we made the simple assumption that all CRs injected by novae travel freely and compute their influence in the surrounding media as an optimistic scenario to study if their influence could be significant. In	this section we will, however, consider a more realistic case derived from recent measurements: recurrent novae bore through	the ISM	to produce cavities in which there is a shock between the material expelled by the continuous eruptions and the ISM, the CRs injected by them are confined in these regions and we will study their effects. The discovery of a cavity around RS Oph \citep{2024MNRAS.529L.175H}, resembling the NSR discovered around M31N 2008-12a \citep{2019Natur.565..460D}, gives rise to the possibility of such a phenomenon around RS Oph. We can thus interpret this cavity as produced by recurrent injections of CRs by the nova in time scales of Myrs, filling a region in the sky with an angular size of $\sim$0.6~deg. We search for gamma rays on this region using data from the Large Area Telescope (LAT) on board the {\it Fermi} Gamma-ray Space Telescope \citep{Atwood_2009} to constrain the content of CRs.
\subsection{Fermi-LAT Analysis}
The gamma-ray emission detected by the LAT during the 2021 eruption declined quickly below the detectable level after $\sim$1~month \citep{Fermi_RSOph}. To evaluate the steady emission due to continuous injection of protons in multiple outbursts, we used the $\sim$15 year dataset covering the period before and after it. %to evaluate the CRs content on the bubble described in \citep{2024MNRAS.529L.175H}.

We performed a binned likelihood analysis using {\it Fermi}-LAT data spanning from MJD 54682.655 to 60377.351 while eliminating time between 59433.5 and 59533.5 corresponding to the recent 2021 outburst of RS Oph. 
RS Oph was not significantly detected 50 days post outburst. We conservatively chose 100 days post outburst to ensure there was no residual contamination in the NSR \citep{Fermi_RSOph}.  
The analysis was performed using \texttt{Fermipy} 1.2.2\footnote{\href{https://fermipy.readthedocs.io/en/latest/}{https://fermipy.readthedocs.io/en/latest/}} and \texttt{Fermitools} v2.2.0\footnote{\href{https://github.com/fermi-lat/Fermitools-conda/}{https://github.com/fermi-lat/Fermitools-conda/}}.
We use {\it Fermi}-LAT data with a $15^\circ$ diameter region of interest (ROI) surrounding RS Oph (RA=267.555$^\circ$, DEC=-6.708$^\circ$) and 4FGL-DR4 as the baseline model to populate source within a $20^\circ$ radius of RS Oph, \texttt{gll\_iem\_v07} as the Galactic diffuse component and \texttt{iso\_P8R3\_SOURCE\_V3\_v1} as the isotropic diffuse component.
The analysis covers an energy range from 100~MeV to 500~GeV divided into 8 energy bins per decade while the spatial binning uses 150 $\times$ 150 bins. 
The joint likelihood analysis was subdivided into the \texttt{PSF0/1/2/3} event types with a zenith angle cut of $90^\circ$ for each event type to eliminate emission from the Earth's Limb. 
While performing the likelihood maximization, the normalization and shape parameters of all sources within a $5^\circ$ radius from RS Oph, including the Galactic and isotropic diffuse components, are left free within the fit. 
% The {\it Fermi}-LAT PSF is on the order of $5^\circ$ but as RS Oph is $\geq 10^\circ$ from the Galactic plane and no source was detected, the results are consistent with freeing sources with larger radius. 

\begin{figure}
	\includegraphics[width=\columnwidth]{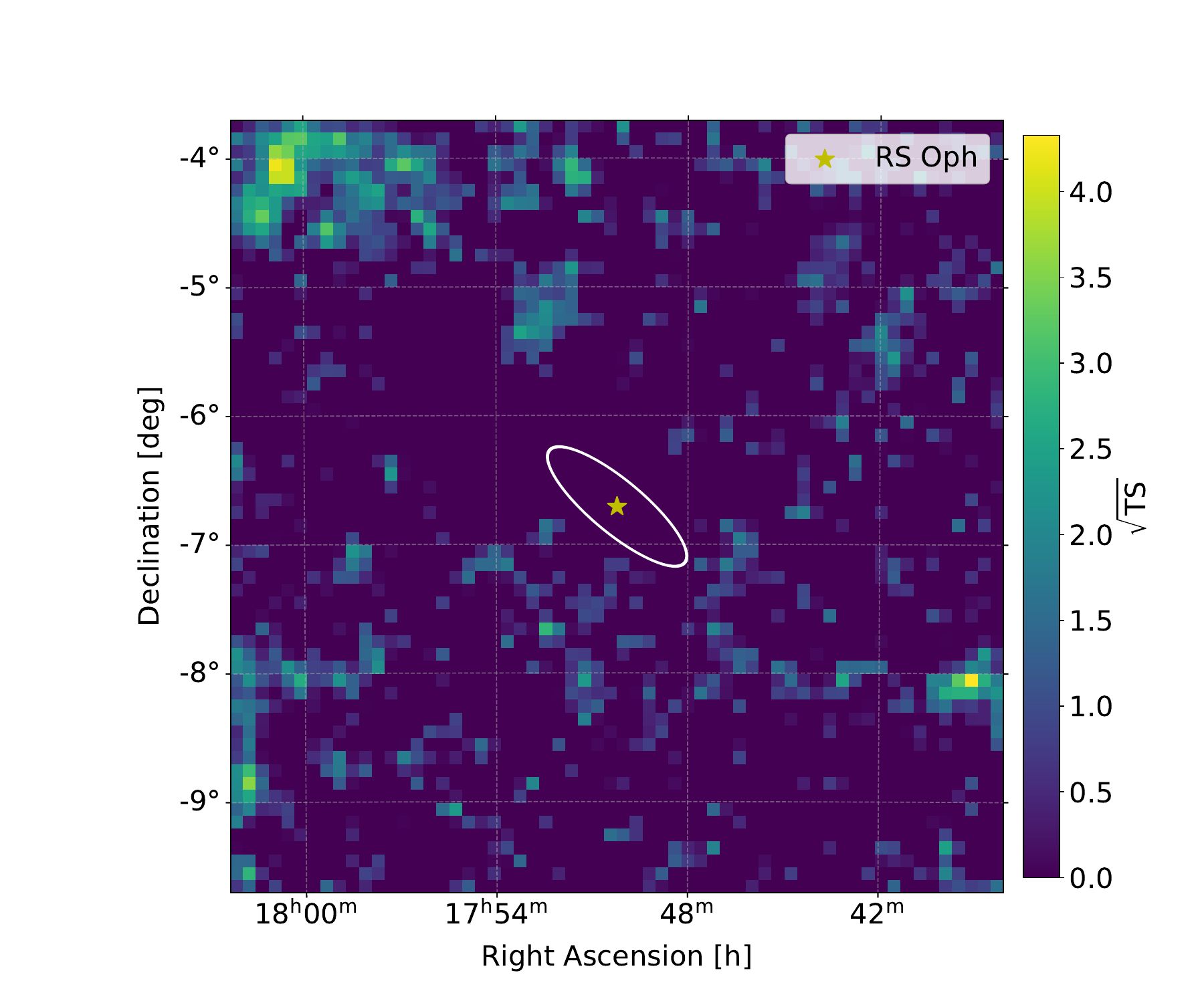}
    \caption{{\it Fermi}-LAT skymap in significance derived from test statistics. In white we can see the template used to derive the $\gamma$-ray upper limits. The square root of the TS, see equation \ref{eqn:TS}, is the significance detecting a new source against the null hypothesis. The parameters for the template are taken from \citet{2024MNRAS.528.3531H}.} 
    \label{fig:fermi_skymap}
\end{figure}

\begin{figure}
	\includegraphics[width=\columnwidth]{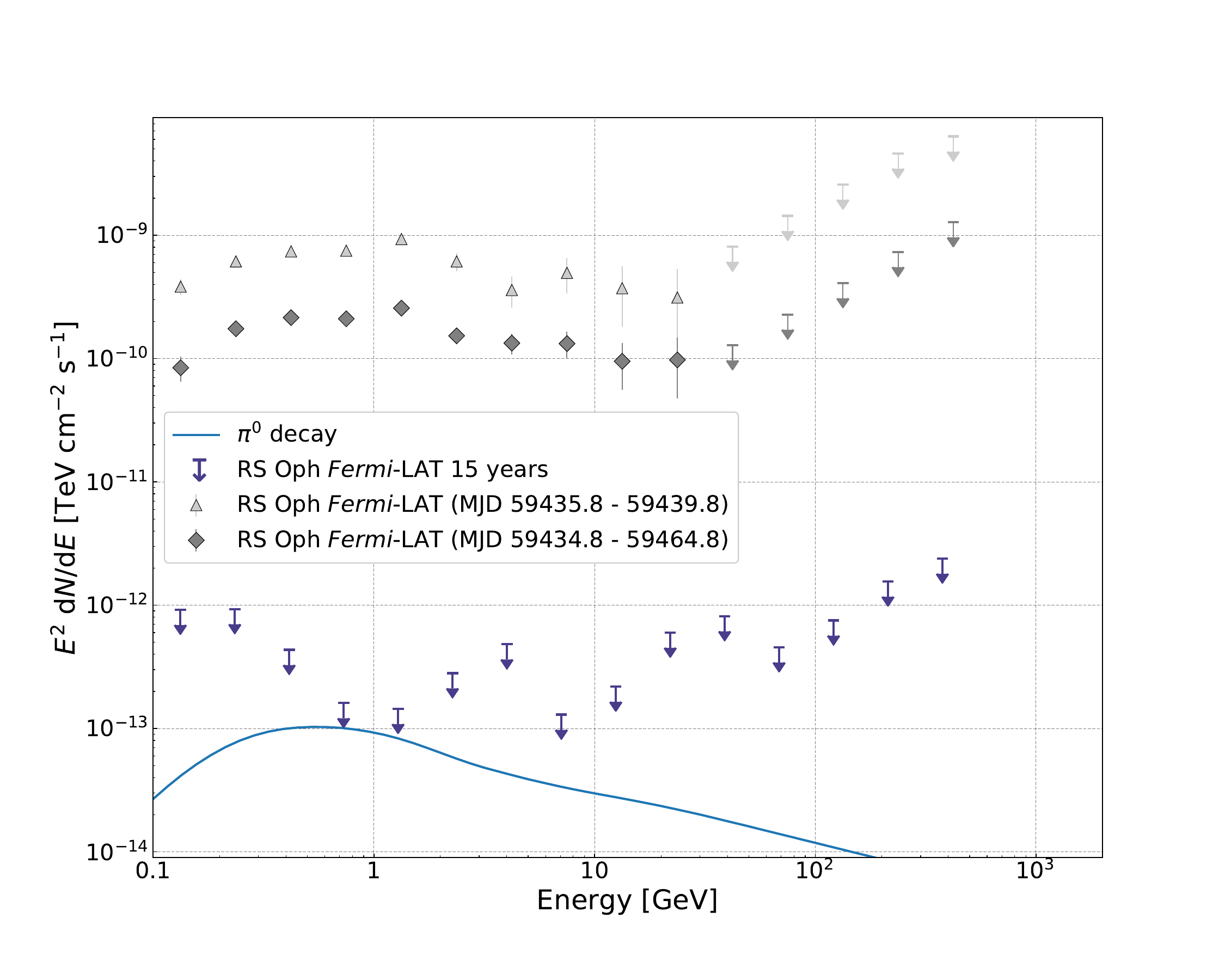}
    \caption{{\it Fermi}-LAT 95\% upper limits on RS Oph 15 year emission using the template from \citet{2024MNRAS.529L.175H} compared with the $\pi^0$ model. The Spectral Energy Distributions from the outburst period of RS Oph are from \citet{MAGIC_RSOph} for the time range shown in the legend where RS Oph is assumed to be a point source.}
    \label{fig:fermi_uls}
\end{figure}

The template we use considers the NSR with a semi-major axis 40~arcmin, an eccentricity of 0.95, and a position angle of 50~deg -the shape extracted in \citep{2024MNRAS.529L.175H}- as it can be seen in Figure \ref{fig:fermi_skymap}.
Notably this is largely consistent with the known bipolar ejecta from RS Oph as demonstrated with later time radio interferometry measurements \citep{2022A&A...666L...6M,2023MNRAS.523..132D}.  
No {\it Fermi}-LAT source was found at the position of RS Oph during the inter-eruption period, and upper limits seen in Fig. \ref{fig:fermi_uls}. 
We take the difference in log-likelihood of the model without the RS Oph NSR (the null hypothesis) to that of the log-likelihood with the RS Oph NSR (test hypothesis).
Using the following equation from \citet{1996ApJ...461..396M}:
\begin{equation}
    \text{TS} \equiv -2 \left( \mathcal{L}_{\text{null}} - \mathcal{L}_{\text{test}}\right),
    \label{eqn:TS}
\end{equation}
where TS is the test statistic, $\mathcal{L}_{\text{null}}$ is the log-likelihood of the null hypothesis, and $\mathcal{L}_{\text{test}}$ is the log-likelihood of the test hypothesis. The significance of detection of a source is defined as the square-root of the TS.  We find a TS = 0.01 for the RS Oph NSR which is well below the TS$\geq$25 threshold to claim a detection\footnote{\href{https://fermi.gsfc.nasa.gov/ssc/data/analysis/documentation/Cicerone/Cicerone_Likelihood/Likelihood_overview.html}{\textit{Fermi}-LAT Likelihood Overview}}. 
The 95\% integrated flux upper limit  for the RS Oph NSR is $3.25 \times 10^{-10}~$cm$^{-2}~$s$^{-1}$, integrated from 100 MeV to 500 GeV.
%Fig. \ref{fig:fermi_uls} was used to constrain the $\pi^0$ emission model. 

% I will fill this on Monday and Tuesday.  
% We analyzed {\it Fermi}-LAT data spanning from MJD xxx to yyy. The analysis was performed using \texttt{fermipy} vxx.x. Events with xxx, yyy....(DESCRIPTION OF THE FERMI ANALYSIS). We used a template that considers the NSR with semi-major axis 40', an ellipticity of 0.4 and a position angle of 45$^\circ$ -the shape extracted in \citep{2024MNRAS.529L.175H}- as it can be seen in Figure \ref{fig:fermi_skymap}.

\subsection{CRs in the RS Oph NSR}
Taking into account a nova energy output used to accelerate protons is $4.4 \times 10^{43}$ erg \citep{MAGIC_RSOph}, and the estimates from \citep{2024MNRAS.529L.175H} of 650,000 eruptions during 25 Myr, we obtain that the total energy into accelerated protons is: 

\begin{equation}
\label{eq:total_energy_novae}
{W_{\rm p} }\sim 3 \times 10^{49} {\rm erg.}
\end{equation}

%We can also consider all the energy produced in the nova explosion, given by equation 11 of \citep{MAGIC_RSOph}, $2 \times 10^{44}$ erg and making the same assumptions as above, calculate an energy output of $\sim 3 \times 10^{50}$ erg.

This quantity may be lower if the energy output that goes into accelerated protons is lower at the initial stages of the lifetime of the nova, so it could be considered an upper limit to the total energy injected into ultra-relativistic particles. To compute the maximum amount of energy allowed by the {\it Fermi}-LAT upper limits, we used \texttt{naima} \citep{2015ICRC...34..922Z} and \texttt{GAMERA} \citep{2015ICRC...34..917H}. We assumed that CR protons interact with the medium, that the total amount of energy injected into accelerated protons is $W_{\rm p}$ and gamma rays are produced via $\pi^0$ decay. The injected proton spectrum is a power-law defined between 100 MeV and 10 TeV following the form:

\begin{equation}
\label{eq:power-law}
\frac{{\rm d}N}{{\rm d}E} = f_0 \left( \frac{E}{E_0} \right)^{-\Gamma}
\end{equation}

with $E_0$=1 TeV, normalization $f_0=2\times10^{46}$ erg$^{-1}$ and spectral index $\Gamma$ = 2.5. This spectrum is directly dependent on the particle density with which the accelerated protons is considered to be interacting, and in this case it is n=1 particle/cm$^3$, of the same order of the ISM. The chosen spectral index and normalization in the particle spectrum are given by the maximum gamma-ray spectrum allowed by the {\it Fermi}-LAT upper limits in the $\sim$GeV energy range and slight modifications that also fit our data in spectral index and normalization are possible.  Given this model, the upper limit on the $\gamma$-ray luminosity above 100 MeV is $L_\gamma < 4\times10^{32}$ erg/s, while the total energy $W$ needed above 100 MeV to be injected into accelerated protons that produce gamma rays compatible with {\it Fermi}-LAT spectrum is:

\begin{equation}
\label{eq:total_energy_fermi}
W = 10^{49} \left(\frac{n}{1\ \mathrm{cm}^{-3}}\right)^{-1} \left(\frac{d}{2.45\ \mathrm{kpc}}\right)^{-2} \mathrm{erg} 
\end{equation}

where $d$ is the distance to the source and $n$ the atomic density of the cavity. If we compare Equations \ref{eq:total_energy_novae} and \ref{eq:total_energy_fermi}, we obtain an upper limit on the ISM atomic density of $n_{d=2.45 \rm{kpc}}<0.4$ cm$^{-3}$, around one order of magnitude smaller than that of the ISM \citep{1998ApJ...506..329W}. If we want to compare the value obtained in \citet{2024MNRAS.529L.175H} with our estimate here, we need to assume a distance of 1.4 kpc as done in that reference.  In this case, the limit derived is $n_{d=1.4 \rm{kpc}}\sim0.1$ cm$^{-3}$, similar to the atomic density calculated in the aforementioned reference.
%For this particular case, not more than 20\% of the total energy employed by the nova to accelerate cosmic rays into the initial stages remain into these particles. This puts a further limit to that established in \cite{MAGIC_RSOph, HESS_RSOph}. 
We note that this estimate depends on the assumption that all nova eruptions produce the same quantity of ultra-relativistic protons and on the assumed distance to the source: closer distances would imply lower atomic densities. The result is also based on the assumption that CRs do not escape from these sources and remain trapped in the NSR bubble. Taking into account that the atomic density derived using our calculations is at the same level that the one inferred in \citet{2024MNRAS.529L.175H}, the only possible explanations are that we are at the level of detection of the source or that the average amount of energy per eruption over the 25 Myr considered is smaller than that of the last eruption, or that CRs actually escape from the source.

% \begin{figure}
% 	\includegraphics[width=\columnwidth]{figures/ULs_RSOph.pdf}
%     \caption{{\it Fermi}-LAT Upper limits on RS Ophiuchi 15 year emission.}
%     \label{fig:fermi_uls}
% \end{figure}

%\section{Discussion}
%It is striking how the maximum energy of accelerated CRs derived by MAGIC is similar to that of local proton kink \citep{PhysRevLett.114.171103, PhysRevLett.122.181102, 2019SciA....5.3793A}

%and the positron curvature \cite{PhysRevLett.122.041102}

%According to the work of \citet{2020PhRvD.101h3018A}, the CR spectrum is the same everywhere in the Galaxy.

\section{Conclusions}

We evaluated the contribution of the TeV gamma-ray novae population to the Galactic CR sea. Even though novae have been proven to be able to accelerate CRs up to multi-TeV energies \citep{MAGIC_RSOph,HESS_RSOph} we derived that a population of RS Oph-like novae, with similar energetic output, would not contribute significantly to the CR sea ($<0.2\%$), neither to large regions around them, that is, local bubbles with radii of $>$few pc.

% \wipeouteow{we established that they are not significant contributors to either the CR sea in the MW ($<0.2\%$) or to local CR bubbles with radii $>$few pc. }

Local hadronic CRs could come from novae only in extreme conditions, with novae needing to be nearby (few tens of parsec) and regular (few decades cadence). Using simple calculations, we conclude that only a non-significant fraction ($<10\%$) of the Galaxy can be covered by nova remnants. These NSRs might be abundant in our Galaxy, however, they are difficult to detect in gamma rays: the continuous outbursts during the nova lifetime blow away the matter of ISM around it, creating a cavity around the binary systems of low-density medium \citep{classical_novae}. This is confirmed by the analysis
of the LAT data on the NSR associated with RS Oph. When excluding the period where the latest outburst was recorded, no gamma-ray source was found on the region defined by the infrared cavity found by \citet{2024MNRAS.529L.175H}. The upper limits allow us to establish an upper limit to the ISM atomic density of $n<0.1$ cm$^{-3}$, under the hypothesis of a history of similar energetic outbursts to the one in 2021. This density is a factor 10 lower than the average density in the ISM and in agreement with the simulations described in \citealt{2024MNRAS.529L.175H}. These cavities are common around SNR explosions and stellar clusters with strong winds and reflect the strong effect of stellar activity on the surrounding ISM up to a hundred-parsec scale. LAT observations from other energetic closeby gamma-ray novae off-outburst can therefore be used to map the ISM densities on different regions in the Galaxy.

%Is this paper relevant for this discussion? https://arxiv.org/pdf/0908.3810
%low density of the ISM in their region  may prevent them from glowing in $\gamma$ rays with a flux that is detectable by current or future HE or VHE $\gamma$-ray instruments. We established a limit on the ISM energy density of $n<0.1$ cm$^{-3}$, much lower than that of the average in the ISM and in agreement with the estimates of \citep{2024MNRAS.529L.175H}.

%Local leptonic CRs at $\sim$hundreds of GeV could only come from novae if the source is closer 

\begin{acknowledgements}
We thankfully acknowledge the comments from the anonymous referee that improved greatly the manuscript. 
R.L.-C., acknowledges the Ram\'on y Cajal program through grant RYC-2020-028639-I, the financial support from the Spanish "Ministerio de Ciencia e Innovaci\'on" through grant PID2022-139117NB-C44, Grant CNS2023-144504 funded by MICIU/AEI/ 10.13039/501100011033 and by the European Union NextGenerationEU/PRTR and the Severo Ochoa program through grant CEX2021-001131-S funded by MCIN/AEI/10.13039/501100011033. He also acknowledges the European Union's Recovery and Resilience Facility-Next Generation, in the framework of the General Invitation of the Spanish Government’s public business entity Red.es to participate in talent attraction and retention programmes within Investment 4 of Component 19 of the Recovery, Transformation and Resilience Plan.

The \textit{Fermi} LAT Collaboration acknowledges generous ongoing support
from a number of agencies and institutes that have supported both the
development and the operation of the LAT as well as scientific data analysis.
These include the National Aeronautics and Space Administration and the
Department of Energy in the United States, the Commissariat \`a l'Energie Atomique
and the Centre National de la Recherche Scientifique / Institut National de Physique
Nucl\'eaire et de Physique des Particules in France, the Agenzia Spaziale Italiana
and the Istituto Nazionale di Fisica Nucleare in Italy, the Ministry of Education,
Culture, Sports, Science and Technology (MEXT), High Energy Accelerator Research
Organization (KEK) and Japan Aerospace Exploration Agency (JAXA) in Japan, and
the K.~A.~Wallenberg Foundation, the Swedish Research Council and the
Swedish National Space Board in Sweden.
 
Additional support for science analysis during the operations phase is gratefully
acknowledged from the Istituto Nazionale di Astrofisica in Italy and the Centre
National d'\'Etudes Spatiales in France. This work performed in part under DOE
Contract DE-AC02-76SF00515.

\end{acknowledgements}

   \bibliographystyle{aa} % style aa.bst
   \bibliography{novae_crs} % your references Yourfile.bib

\end{document}